\documentclass[reprint,
 amsmath,amssymb,
 aps,
]{revtex4-1}

\usepackage{graphicx}
\usepackage{dcolumn}
\usepackage{bm}
\usepackage{tensor} 

\begin{document}

\title{On (non-)dynamical dark energy}
\author{Winfried Zimdahl$^{1}$}
 
\author{J\'ulio Fabris$^{1,2}$}%
\author{Hermano Velten$^{3}$}%
 \altaffiliation[]{hermano.velten@ufop.edu.br (corresponding author)}
\author{Ram\'on Herrera$^{4}$}%
\affiliation{%
$^{1}$N\'ucleo COSMO-UFES \& Departamento de F\'isica,  Universidade Federal do Esp\'irito Santo (UFES)\\
 Av. Fernando Ferrari s/n CEP 29.075-910, Vit\'oria, ES, Brazil }%
\affiliation{%
$^{2}$National Research Nuclear University MEPhI, Kashirskoe sh. 31, Moscow 115409, Russia}%
\affiliation{%
$^{3}$Departamento de F\'isica, Universidade Federal de Ouro Preto (UFOP), Campus Universit\'ario Morro do Cruzeiro, 35.400-000, Ouro Preto, Brazil}%
\affiliation{%
$^{4}$Instituto  de  F\'isica,  Pontificia  Universidad  Cat\'olica de  Valpara\'iso,  Casilla  4059,  Valpara\'iso,  Chile.}%
\date{\today}

\begin{abstract}The current Universe is composed by a mixture of relativistic species, baryonic matter, dark matter and dark energy which evolve in a non-trivial way at perturbative level. An advanced description of the cosmological dynamics should include non-standard features beyond the simplistic approach idealized by the standard cosmology in which cosmic components do not interact, are adiabatic and dissipationless.  We promote a full perturbative analysis of linear scalar perturbations of a non-interacting cosmological model containing baryons, dark matter (both pressurless) and a scalar field allowing for the presence of relative entropic perturbations between the three fluids. Assuming an effective scalar-field sound speed equal to one and neglecting anisotropic stresses we establish a new set of equations for the scalar cosmological perturbations. As a consequence of this new approach, we show that tiny departures from a constant scalar field equation of state $w_{S}=-1$ damage structure formation in a non-acceptable manner. Hence, by strongly constraining $w_S$ our results provide compelling evidence in favor of the standard cosmological model and rule out a large class of dynamical dark energy models.
\begin{description}
\item[PACS numbers: 98.80.-k, 95.36.+x, 04.25.Nx]
\end{description}
\end{abstract}

\maketitle

{\bf Introduction - } The dynamics of matter fluctuations in the expanding Universe is the key element in theories of cosmic structure formation. According to the cosmological standard model, a matter-dominated phase of sufficient duration is required to account for the growth of initially small inhomogeneities in the early Universe into the currently observed highly nonlinear matter agglomerations. Indeed, the electromagnetically observed matter distribution in galaxy catalogues describes the distribution of baryonic matter.
The growth rate of baryonic matter perturbations depends on the composition of the cosmic substratum.
For successful structure formation it is a crucial feature that after recombination baryonic matter perturbations fall into the potential  wells of dark-matter (DM) fluctuations.
As soon as some form of dark energy (DE) comes to dominate the cosmic dynamics the impact of this component on the baryon fluid has to be considered. The details of this impact depend on how the unknown DE component is modeled.
In standard $\Lambda$CDM cosmology the baryonic matter growth (driven by the DM overdensitites) is attenuated through the modified (due to the existence of a non-vanishing cosmological constant) homogeneous and isotropic background dynamics. By definition, $\Lambda$ is a constant, i.e., it behaves dynamically as a homogeneous fluid.
The situation is different in dynamical DE models. Here, inhomogeneities appear quite naturally and such kind of DE is intrinsically fluctuating. The problem then is, how  much potential DE inhomogeneities contribute to the total energy-density perturbations of the cosmic medium, equivalent to their influence on the gravitational potential, and how much the baryonic matter growth rate is influenced by the presence of DE perturbations. Since different DE models predict a different influence on the perturbation dynamics, in particular also a different influence on the growth rate of baryonic matter fluctuations, observational growth data may be used to discriminate between competing dark-energy models.
In this letter we model the cosmic medium as a mixture of three separately conserved perfect fluids, one of them being
pressureless baryonic matter.
The remaining two components represent different fluid configurations of the cosmological dark sector.
The first one describes a generic component with vanishing sound speed, the second one a general fluid component with a sound speed equal to the speed of light.
For our specific analysis the first dark-sector component behaves in the background as a fluid with constant equation of state (EoS) parameter,
the second one is the fluid representation of a scalar field (SF) (quintessence or phantom) in  CPL parametrization \cite{CPL}. 
This choice is sufficiently flexible to investigate deviations from the standard model in different directions.
A three-fluid model is generally equipped with three four-velocities which, in the general case, differ from each other at the perturbative level, although in the homogeneous and isotropic standard-model background they are assumed to coincide.
With the help of the combined energy-momentum conservation equations for the components we clarify whether or not these differences are relevant.

Our study differs from previous investigations by the manner in which the pressure perturbations of the cosmic medium are determined. An appropriate modeling of pressure perturbations beyond the purely adiabatic contributions is crucial in order to close the system of perturbation equations.
The usual procedure to account for non-adiabatic perturbations is to assume an effective sound speed of the DE component as an additional parameter which, together with an EoS parameter for the homogeneous and isotropic background, fixes the cosmological dynamics \cite{Hu:1998kj}.
It turned out that this parameter is difficult to constrain from available observational data \cite{DeDeo:2003te, Bean:2003fb, Hannestad:2005ak, Mehrabi:2015hva}.
Alternatively, the use of the effective field theory approach for dark energy \cite{Gubitosi:2012hu} allows for an elegant exploration of the perturbative features detached from the background expansion. One can also phenomenologically parametrize the two independent functions of the gravitational potentials \cite{Amendola:2016saw}. 
Results along these lines have been recently reported by the Planck team \cite{Ade:2015rim,Aghanim:2018eyx} and mild deviations from the standard cosmology are found. 

Our focus here is on the circumstance that in a multi-component medium there are not just intrinsic non-adiabatic perturbations, modeled by an effective sound speed of each of the components, but there appear non-adiabatic relative perturbations which additionally contribute to the total pressure perturbation of the cosmic fluid even if each of the individual components is adiabatic. (see \cite{Dent:2008ek, Velten:2017mtr, Velten:2018bws}).
This also means, the intrinsic non-adiabatic contributions do not simply add up to the total non-adiabatic pressure perturbation.
The novel feature of our approach consists in taking into account the entire set of non-adiabatic perturbations in a three-component cosmic medium and to
explore the consequences for the growth rate of matter perturbations.
To describe the linear (subhorizon) perturbation dynamics we establish a coupled system of equations for the total energy-density perturbations (equivalent to a second-order equation for the gravitational potential) and two first-order equations for two relative perturbation variables.
On this basis we shall check to what extent the observed $f\sigma_8$ data tolerate small deviations from the standard $\Lambda$CDM model.

While the interplay between adiabatic
and non-adiabatic perturbations on superhorizon scales is well understood \cite{Wands:2000, Weinberg:2003sw},
its role for late-time structure formation on scales well within the horizon does not seem to have adequately studied so far.

{\bf Modeling the system} - 
The cosmic medium is modeled as a perfect fluid with the energy-momentum tensor
\begin{equation}
T_{ik} = \rho u_{i}u_{k} + p h_{ik}\ \,, \qquad T_{\ ;k}^{ik} = 0\,.
\label{T}
\end{equation}
Here, $\rho = T_{ik}u^{i}u^{k}$ is the total energy density and $p = \frac{1}{3}T_{ik}h^{ik}$ is the total  pressure with $ h^{ik} = g^{ik} + u^{i} u^{k}$. The four-velocity $u^{i}$ of the cosmic medium is normalized to $u^{i}u_{i} = -1$.
The expansion scalar $\Theta \equiv u^{a}_{;a}$ is determined by the Raychaudhuri equation which, in the absence of shear and rotation,
reduces to
\begin{equation}
\dot{\Theta} + \frac{1}{3}\Theta^{2} - \dot{u}^{a}_{;a} + 4\pi G \left(\rho + 3
p\right) = 0, \label{Ray}
\end{equation}
with $\dot{u}^{a}= u^{a}_{;n}u^{n}$ being the fluid acceleration.
The total energy-momentum tensor in (\ref{T}) is split into a DM component (subindex D), a SF component (subindex S) and a baryonic component (subindex B), i.e.,
$T^{ik} = T_{D}^{ik} + T_{S}^{ik}  + T_{B}^{ik}$. 
Each of these components is supposed to have a separately conserved perfect-fluid structure as well, i.e., $T_{M\ ;k}^{ik} = 0$ with $M=D, S, B$.
In general, the four-velocities $u^{i}_{M}$ of the components are different from each other and from the total
four-velocity $u^{i}$.
DM is characterized by an
EoS parameter $w_{D}\equiv p_{D} / \rho_{D}$ where $w_{D}$ is a constant.
For the SF we use an effective fluid description with an effective EoS parameter $w_{S}\equiv p_{S} / \rho_{S} $ which is supposed to cover both quintessence and phantom configurations. Finally, the baryons are modeled  as pressureless fluid with an EoS $p_B=0$.

In the spatially homogeneous and isotropic background
all four-velocities
are assumed to coincide:
$u_{M}^{a} = u^{a}$.
Under this condition the energy-conservation equation for the DM can be integrated
\begin{equation}\label{}
\rho_{D} = \rho_{D0} \,a^{-3\left(1+w_{D}\right)},
\end{equation}
where $a$ is the scale factor of the Robertson-Walker metric with its present value $a_{0}$ normalized to $a_{0}=1$.
For the fluid representation of the scalar field we use the
CPL parametrization $w_{S} = w_{0} + w_{1}\left(1-a\right)$ with the help of which the energy-conservation equation can be solved explicitly:
\begin{equation}\label{}
\rho_{S} = \rho_{S0}a^{-3\left(1 + w_{0} + w_{1}\right)}\exp{\left[3w_{1}\left(a-1\right)\right]}.
\end{equation}
Indeed it is always possible to find a scalar field potential which reproduces a given cosmological expansion \cite{Padmanabhan:2002cp}.
For the baryons we have $p_{B} = 0$ and $\rho_{B} = \rho_{B0} a^{-3}$.
With the explicitly known energy densities of the components the Hubble rate is determined by Friedmann's equation
$3 H^{2} = 8\pi G \rho = 8\pi G \left(\rho_{D} + \rho_{S} + \rho_{B}\right).$

{\bf Cosmological perturbations - } Scalar metric perturbations are described by the line element (in the longitudinal gauge)
\begin{equation}
\mbox{d}s^{2} = - \left(1 + 2 \phi\right)\mbox{d}t^2 +
a^2\left(1-2\psi\right)\delta _{\alpha \beta} \mbox{d}x^\alpha\mbox{d}x^\beta \ .\label{ds}
\end{equation}
The perturbed time components of the four-velocities are (first-order fluid quantities will be denoted by a hat symbol)
$\hat{u}_{0} = \hat{u}^{0} = \hat{u}_{M}^{0} =\hat{u}_{M0}  = \frac{1}{2}\hat{g}_{00} = - \phi.$
Our restriction to scalar perturbations implies that the spatial components of the four velocity are gradients of a scalar: $\hat{u}_{\alpha} = v_{,\alpha}$.
Likewise, the scalar parts of the spatial components of the components four-velocities are introduced through $\hat{u}_{M\alpha} = v_{M,\alpha}$.
The definition of $\Theta$ yields  
\begin{equation}
\label{}
\hat{\Theta} = -\frac{k^{2} v}{a^2}  -
3\dot{\psi} - 3 H\phi,
\end{equation}
with similar expressions for the other components.
Here, $k$ is the comoving wave number.
At linear order the components of the energy-momentum tensors simply add which also establishes a relation between the perturbations of the four-velocities.
The comoving first-order fractional energy-density perturbation $\delta^{c} \equiv \frac{\hat{\rho}^{c}}{\rho}$, where
$\delta^{c} \equiv \delta + \frac{\dot{\rho}}{\rho}v$ and $\delta \equiv \frac{\hat{\rho}}{\rho}$,
obeys
the second-order equation
\begin{eqnarray}
&&\delta^{c\prime\prime} + \left[\frac{3}{2}-\frac{15}{2}\frac{p}{\rho}+ 3\frac{p^{\prime}}{\rho^{\prime}}\right]\frac{\delta^{c\prime}}{a} \nonumber \\
&&- \left[\frac{3}{2} + 12\frac{p}{\rho} - \frac{9}{2}\frac{p^{2}}{\rho^{2}} - 9\frac{p^{\prime}}{\rho^{\prime}}
\right]\frac{\delta^{c}}{a^{2}}
+ \frac{k^{2}}{a^{2}H^{2}}\frac{\hat{p}^{c}}{\rho a^{2}}
= 0,
  \label{dddeltak}
\end{eqnarray}
where $\hat{p}^{c} \equiv \hat{p} + \dot{p}v$ denotes the comoving total pressure perturbations. The symbol prime $ ``\,^{\prime}\,"$ means a derivative with respect to the scale factor.
Non-vanishing pressure perturbations are a characteristic feature  of dynamical dark energy.  In the standard $\Lambda$CDM model there are no pressure perturbations.
Equation (\ref{dddeltak}) is the result of the energy-momentum conservation equations combined with the Raychaudhuri equation.
The gauge-invariant quantities (superscript c) are defined with respect to the total four-velocity of the cosmic medium.
Since there is no EoS of the type $p = p(\rho)$ for the medium as a whole, there appears a coupling
to the dynamics of the components via the (gauge-invariant) comoving pressure perturbation $\hat{p}^{c}$.
If there were a relation $p=p(\rho)$, Eq.~(\ref{dddeltak}) would be a closed equation for $\delta^{c}$. The absence of such relation is equivalent to the existence of non-adiabatic perturbations.
Notice that it is this quantity $\delta^{c}$ which appears in the Poisson-type equation for the gravitational potential,
\begin{equation}
\label{}
-\frac{2}{3}\frac{k^{2}}{H^{2}a^{2}}\psi = \delta^{c}.
\end{equation}
Since we assumed a perfect-fluid structure of the medium, there are no anisotropic stresses and $\psi = \phi$.

{\bf Introducing non-adiabatic perturbations -} 
If the total medium were adiabatic, the pressure perturbations would be given by
$\hat{p} = \frac{\dot{p}}{\dot{\rho}}\hat{\rho}$.
The difference $\hat{p} - \frac{\dot{p}}{\dot{\rho}}\hat{\rho}$ characterizes deviations from adiabaticity.
The total  non-adiabatic pressure perturbation can be written in terms of the contributions of the components.
The (gauge-invariant) non-adiabatic parts of the component are
\begin{equation}\label{}
\hat{p}_{Snad} \equiv \hat{p}_{S} - \frac{\dot{p}_{S}}{\dot{\rho}_{S}}\hat{\rho}_{S} \,,\qquad
\hat{p}_{Dnad} \equiv \hat{p}_{D} - \frac{\dot{p}_{D}}{\dot{\rho}_{D}}\hat{\rho}_{D}.
\end{equation}
It is convenient to define the (gauge-invariant) relative perturbations
\begin{equation}\label{Relpert}
S_{BD} \equiv \delta_{B} - \frac{\delta_{D}}{1+w_{D}}, \quad \delta_{B}\equiv \frac{\hat{\rho}_{B}}{\rho_{B}},\quad
\delta_{D}\equiv \frac{\hat{\rho}_{D}}{\rho_{D}}
\end{equation}
and
\begin{equation}\label{Relpert2}
S_{BS} \equiv \delta_{B} - \frac{\delta_{S}}{1+w_{S}},\quad
\delta_{S}\equiv \frac{\hat{\rho}_{S}}{\rho_{S}}.
\end{equation}
With the help of the fractional quantities $\Omega_{B} \equiv \rho_{B}/\rho, \, \Omega_{D} \equiv \rho_{D}/\rho, \, \Omega_{S} \equiv \rho_{S}/\rho$ the comoving pressure perturbation becomes
\begin{eqnarray}
 \label{hatp/rho}
&&\frac{\hat{p}^{c}}{\rho} = \frac{p^{\prime}}{\rho^{\prime}}\delta^{c}
+ \left(1 + w_{D}\right)\Omega_{D} P_{Dnad} + \left(1 + w_{S}\right)\Omega_{S} P_{Snad}
\nonumber \\
   &&\qquad + \frac{1+w_{S}}{1+w}\Omega_{S}U S_{BS} + \frac{1+w_{D}}{1+w}\Omega_{D} V S_{BD},
 \end{eqnarray}
 where
    \begin{equation}\label{}
   P_{Dnad} \equiv \frac{\hat{p}_{Dnad}}{\rho\left(1 + w_{D}\right)\Omega_{D}},\quad
   P_{Snad} \equiv \frac{\hat{p}_{Snad}}{\rho\left(1 + w_{S}\right)\Omega_{S}}.
 \end{equation}
 The quantities $U$ and $V$ are combinations of background quantities:  
  \begin{equation}\label{}
  U\equiv \frac{p_{D}^{\prime}}{\rho_{D}^{\prime}}\left(1+w_{D}\right)\Omega_{D}
- \frac{p_{S}^{\prime}}{\rho_{S}^{\prime}}\left(\Omega_{B}+\left(1+w_{D}\right)\Omega_{D}\right)
 \end{equation}
   and 
    \begin{equation}\label{}
 V\equiv \frac{p_{S}^{\prime}}{\rho_{S}^{\prime}}\left(1+w_{S}\right)\Omega_{S} - \frac{p_{D}^{\prime}}{\rho_{D}^{\prime}}\left(\Omega_{B}+\left(1+w_{S}\right)\Omega_{S}\right),
 \end{equation}
respectively. Further we have 
 $w = \frac{p}{\rho} = w_{D}\Omega_{D} + w_{S}\Omega_{S}$
 and
 \begin{equation}\label{}
\frac{p^{\prime}}{\rho^{\prime}} = \frac{p_{D}^{\prime}}{\rho_{D}^{\prime}}
\frac{1+w_{D}}{1+w}\Omega_{D}
+\frac{p_{S}^{\prime}}{\rho_{S}^{\prime}}
\frac{1+w_{S}}{1+w}\Omega_{S}.
 \end{equation}
Through  relation (\ref{hatp/rho}) the dynamics of $\delta^{c}$ in (\ref{dddeltak}) is coupled to the relative perturbations
 $S_{BS}$ and $S_{BD}$. Notice that an explicit contribution $S_{DS}$ does not appear in (\ref{hatp/rho}). Of the three variables $S_{BS}$, $S_{BD}$ and $S_{DS}$ only two are independent.
 We consider $\delta^{c}$, $S_{BS}$ and $S_{BD}$ to be the basic variables of our perturbation analysis.
 Our aim will be to establish a closed system of equations for these variables.
So far, the perturbation dynamics (neglecting temporarily our background specifications) is valid for any three-fluid system with one of the components being pressureless. A concrete model is obtained by specifying the pressure perturbations of the components. This has to be done in the rest frames of the components, which, in general, do not coincide with each other and also not with the global rest frame.
The individual comoving pressure perturbations are
$\hat{p}_{D}^{c_{_{D}}} \equiv \hat{p}_{D} + \dot{p}_{D}v_{D}$ for the DM and $\hat{p}_{S}^{c_{_{S}}} \equiv \hat{p}_{S} + \dot{p}_{S}v_{S}$ for the SF.
The corresponding rest-frame energy densities are
 \begin{equation}\label{}
\hat{\rho}_{D}^{c_{_{D}}} 
= \hat{\rho}_{D}^{c}- \dot{\rho}_{D}\left(v - v_{D}\right),
 \end{equation}
for the DM and
 \begin{equation}\label{}
\hat{\rho}_{S}^{c_{_{S}}}  =  \hat{\rho}_{S}^{c} - \dot{\rho}_{S}\left(v - v_{S}\right),
 \end{equation}
for the SF.
Generally, the energy-densities of the components in their own rest-frame do not coincide with the energy-densities of these components measured in the rest frame of the cosmic fluid as a whole.
Obviously, information about the velocity differences is required.

{\bf Dark-sector pressure perturbations -} 
Any phenomenological description of fluid perturbations needs information about the sound speed of the medium.
On physical grounds the sound speed lies between zero and one.
To specify the model we firstly assume the DM to be pressureless, i.e., $\hat{p}_{D}^{c_{_{D}}} = 0$.
The DM pressure perturbation in its own rest frame is zero.
This corresponds to a vanishing effective sound speed of this component. It follows that 
\begin{equation}
\label{}
P_{Dnad} = - \frac{\dot{p}_{D}}{\dot{\rho}_{D}}\frac{\hat{\rho}_{D}^{c_{_{D}}}}{\rho_{D} + p_{D}}.
\end{equation}
Secondly, for the SF we assume $\hat{p}_{S}^{c_{_{S}}} = \hat{\rho}_{S}^{c_{_{S}}}$ then
\begin{equation}
\label{}
P_{Snad} = \left(1 - \frac{\dot{p}_{S}}{\dot{\rho}_{S}} \right)\frac{\hat{\rho}_{S}^{c_{_{S}}}}{\rho_{S} + p_{S}}.
\end{equation}
The SF pressure perturbation in the SF rest frame equals the SF energy-density perturbation in this frame.
This represents the opposite case to that of the DM. Here the effective sound speed is equal to 1.
Our DM-SF model is fully specified by the pressure perturbations $P_{Dnad}$ and $P_{Snad}$ together with the background equations of state $w_{D}$ and $w_{S}$.
The requirement $\hat{p}_{D}^{c_{_{D}}}  = 0$ implies that the velocity potentials $v_{B}$ and $v_{D}$ can be identified.
To determine $P_{Dnad}$ and $P_{Snad}$ we have to know the energy-density perturbations in the rest frames of the components in terms of $\delta^{c}$, $S_{BD}$ and $S_{BS}$.
Use of the conservation equations reveals for the difference of the velocity potentials
\begin{equation}\label{}
v_{D} - v_{S}  = \frac{a^{2}H^{2}}{k^{2}}\cdot Y \cdot H^{-1}
 \end{equation}
 where
 \begin{equation}\label{}
 Y \equiv 
 \frac{a S_{BS}^{\prime} - 3\left(1 - \frac{p_{S}^{\prime}}{\rho_{S}^{\prime}}\right)D_{S}^{c}}{1 + 9\frac{a^{2}H^{2}}{k^{2}}\left(1 - \frac{p_{S}^{\prime}}{\rho_{S}^{\prime}}\right)
\frac{\left(1+w_{D}\right)\Omega_{D} + \Omega_{B}}{1+w}}
\end{equation}
with 
\begin{eqnarray}
D_{S}^{c}&\equiv&\frac{\delta^{c}}{1+w}
 - \frac{\left(1+w_{D}\right)\Omega_{D} + \Omega_{B}}{1+w}S_{BS}
 \nonumber\\&& \qquad\qquad+ \frac{\left(1+w_{D}\right)\Omega_{D}}{1+w}S_{BD}.
\end{eqnarray}
For perturbations well below the horizon scale $\frac{k^{2}}{a^{2}H^{2}} \gg 1$ is valid. Under this condition all relations simplify considerably since the velocity-difference terms  become negligible. On these scales it is no longer necessary to discriminate between the different rest frames. 
For scales of interest here  which are still linear until  today  and at the same time big enough to represent typical large scale structures say, of the order of $k \approx 0.1 h {\mathrm{Mpc}^{-1}}$,  the present value of the factor $k^{2} / a^{2} H^{2}$ is (restoring the units appropriately) $ k^{2}c^{2} / H_{0}^{2} \approx 1.8\cdot 10^{5}$.  Under these conditions a closed system for $S_{BD}$, $S_{BS}$ and $\delta^{c}$ can be derived.
For $S_{BD}$ and $S_{BS}$ we have
\begin{eqnarray}\label{SBCpr}
&&S_{BD}^{\prime} - 3\frac{p_{D}^{\prime}}{\rho_{D}^{\prime}}
\frac{\Omega_{B}+\left(1+w_{S}\right)\Omega_{S}}{1+w}\frac{S_{BD}}{a}
= \nonumber \\
&&\qquad-3 \frac{p_{D}^{\prime}}{\rho_{D}^{\prime}}\frac{1}{1+w}
\left[\frac{\delta^{c}}{a} + \left(1+w_{S}\right)\frac{S_{BS}}{a}\right]
\end{eqnarray}
and
\begin{eqnarray}\label{SBSpr}
&&S_{BS}^{\prime} + 3\left(1 - \frac{p_{S}^{\prime}}{\rho_{S}^{\prime}}\right)\frac{\Omega_{B}+\left(1+w_{D}\right)\Omega_{D}}{1+w}\frac{S_{BS}}{a}
= \nonumber \\
&&\qquad
3\left(1 - \frac{p_{S}^{\prime}}{\rho_{S}^{\prime}}\right)\frac{1}{1+w}
\left[\frac{\delta^{c}}{a} + \left(1+w_{D}\right)\frac{S_{BD}}{a}\right],
\end{eqnarray}
respectively.
Equations (\ref{SBCpr}) and (\ref{SBSpr}) are first-order equations for $S_{BD}$ and $S_{BS}$, respectively, in which
$\delta^{c}$ and $S_{BS}$ (or $S_{BD}$, respectively) appear as inhomogeneities.
The total comoving pressure perturbation  reduces to
\begin{equation}
\frac{\hat{p}^{c}}{\rho} =\frac{1+w_{S}}{1+w}\Omega_{S}\left(\delta^{c} + \Sigma \right),
\label{hatpfin}
\end{equation}
with
\begin{equation}\label{Pscalar++}
\Sigma \equiv \left(1+w_{D}\right)\Omega_{D}S_{BD}
 - \left(\Omega_{B}+\left(1+w_{D}\right)\Omega_{D}\right)S_{BS},
\end{equation}
i.e., they are entirely given in terms of the energy-density perturbations and the relative entropy perturbations.

There are some similarities between our setup and the parametrization of perturbations via equations of state in  \cite{Battye:2013aaa,Battye:2014xna} on the basis of a rather general scalar-field Lagrangian.
In this approach, equations of state for the
entropy perturbation in terms of perturbations of the density, the velocity and
the metric perturbations are introduced in order to obtain closed perturbation equations.
What is different among others is the choice of basic variables. In particular, the entropy perturbations are basic dynamical quantities in our context and neither velocity components nor metric functions do appear explicitly
in the system of equations.

With the pressure perturbation (\ref{hatpfin}) in terms of $\delta^{c}$, $S_{BD}$ and $S_{BS}$, equation (\ref{dddeltak})  for $\delta^{c}$ becomes explicitly
\begin{eqnarray}
&&\delta^{c\prime\prime} + \left[\frac{3}{2}-\frac{15}{2}w+ 3\frac{p^{\prime}}{\rho^{\prime}}\right]\frac{\delta^{c\prime}}{a} - \left[\frac{3}{2} + 12w - \frac{9}{2}w^{2} - 9\frac{p^{\prime}}{\rho^{\prime}}\right.\nonumber \\
&&\left.
\quad - \frac{k^{2}}{a^{2}H^{2}}\frac{1+w_{S}}{1+w}\Omega_{S}\right]\frac{\delta^{c}}{a^{2}} +
\frac{k^{2}}{a^{2}H^{2}}\frac{1+w_{S}}{1+w}\Omega_{S}\Sigma = 0.
  \label{dddeltaf}
\end{eqnarray}
The relative perturbations $S_{BC}$ and $S_{BS}$ appear as inhomogeneities in the equation for $\delta^{c}$.
This equation has to be solved together with (\ref{SBCpr}) and (\ref{SBSpr}).
The set of equations (\ref{SBCpr}), (\ref{SBSpr}) and (\ref{dddeltaf}) represents a closed system for
$\delta^{c}$, $S_{BD}$ and $S_{BS}$.
The baryonic matter perturbations can be obtained as the difference $\hat{\rho}_{B} = \hat{\rho} - \hat{\rho}_{D} - \hat{\rho}_{S}$.
The result is
\begin{equation}\label{deltaB}
\delta_{B}^{c} = \frac{\delta^{c} + \left(1+w_{D}\right)\Omega_{D}S_{BD} + \left(1+w_{S}\right)\Omega_{S}S_{BS}}{1+w_{D}\Omega_{D}  + w_{S}\Omega_{S}}.
\end{equation}
It is this quantity which through the solution of the coupled system (\ref{SBCpr}), (\ref{SBSpr}) and (\ref{dddeltaf})
for $\delta^{c}$, $S_{BD}$ and $S_{BS}$ determines the growth rate of baryonic matter perturbations. The results presented below are obtained from the numerical solution of this set of equations.

We run Eqs. (\ref{SBCpr}), (\ref{SBSpr}) and (\ref{dddeltaf}) with initial conditions for density fluctuation as given by the CAMB code \cite{Lewis:1999bs} in a standard PLANCK values cosmology at high redshifts.  It is worth noting this scale enters horizon during the radiation dominated epoch. Hence, since our equations are designed to the late time cosmology there is no horizon crossing along the numerical evolution. This implies to assume that the initial perturbations are almost adiabatic, i.e., $S_{BD}\approx 0$ and $S_{BS}\approx 0$.
The mere multi-fluid structure of the medium 
requires that there is always a small, non-vanishing admixture of non-adiabaticity. Numerically, we found no difference between the runs for $S_{BD}$ and $S_{BS}$ very small and zero exactly.  
Our figures are obtained with purely adiabatic initial conditions.  
While on superhorizon scales purely adiabatic modes cannot give rise to entropy perturbations \cite{Wands:2000}, our present subhorizon dynamics is different. 

\begin{figure}
\includegraphics[width=\columnwidth]{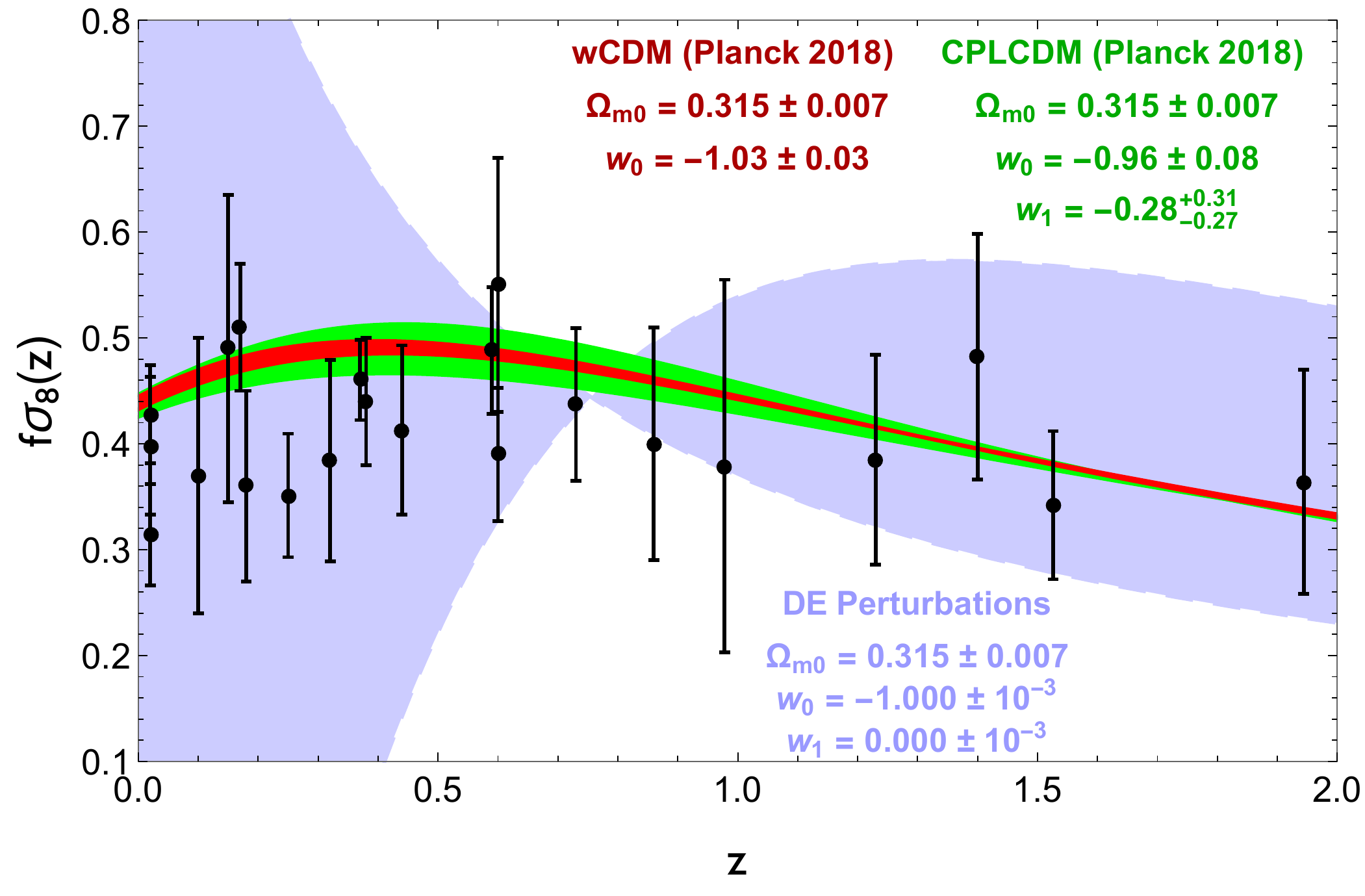}
\caption{Impact of DE perturbations on $f\sigma_8(z)$. The light blue region visualizes the departure from  the standard wCDM cosmology (red) with Planck 2018 parameter values if small deviations of the order of $\pm 10^{-3}$ around $w_{S}=-1$ are admitted. Each colored stripe is filled with all $f \sigma_8(z)$ curves produced with corresponding colored parameter intervals shown in the insets of this figure.
For comparison, the Planck 2018 results for a CPLCDM model without non-adiabatic DE perturbations (green) is also included. 
All cases assume a standard pressureless DM component i.e., $w_{D}=0$. 
At high redshift, all curves show the same asymptotic behavior.}
\label{Fig2}
\end{figure}
In order to provide a measurable quantity to test the evolution of matter perturbations calculated from Eq. (\ref{deltaB}) we compute the growth rate $f(a) =$ d ln$\delta_B$/d ln$a$ together with the variance of the matter density distribution smoothed on spheres of radius $8 h^{-1}$ Mpc, $\sigma_8 (a) = \sigma_{8_0} \, \delta_B (a)/ \delta_B (a=1)$. We have consistently checked that $\delta_B$ indeed follows the evolution of the total matter (baryons + DM) overdensity $\delta_m$ as in the standard cosmological scenario. 
A recent reliable data compilation of the product $f\sigma_8(a)$ - a bias-free cosmological observable - is found in \cite{Sagredo:2018ahx} and is shown in Fig.\ref{Fig2}. 
The light blue region is obtained from  Eq. (\ref{deltaB}) after solving the set (\ref{SBCpr}), (\ref{SBSpr}) and (\ref{dddeltaf}). Its boundaries denote the limits for deviations of the order of  $\pm 10^{-3}$ (for both $w_0$ and $w_1$) from $w_{S}=-1$. 
The predictions for a $wCDM$ standard cosmology are represented by the red stripe ($\Omega_{m0}$ and $w_0 = -1.03\pm 0.03$ \cite{Aghanim:2018eyx}).
In green we also include the Planck 2018 results for a CPLCDM model in which non-adiabatic DE perturbations are neglected. 
The same $\Omega_{m0}$ value was used for all the models.

\begin{figure}
\includegraphics[width=\columnwidth]{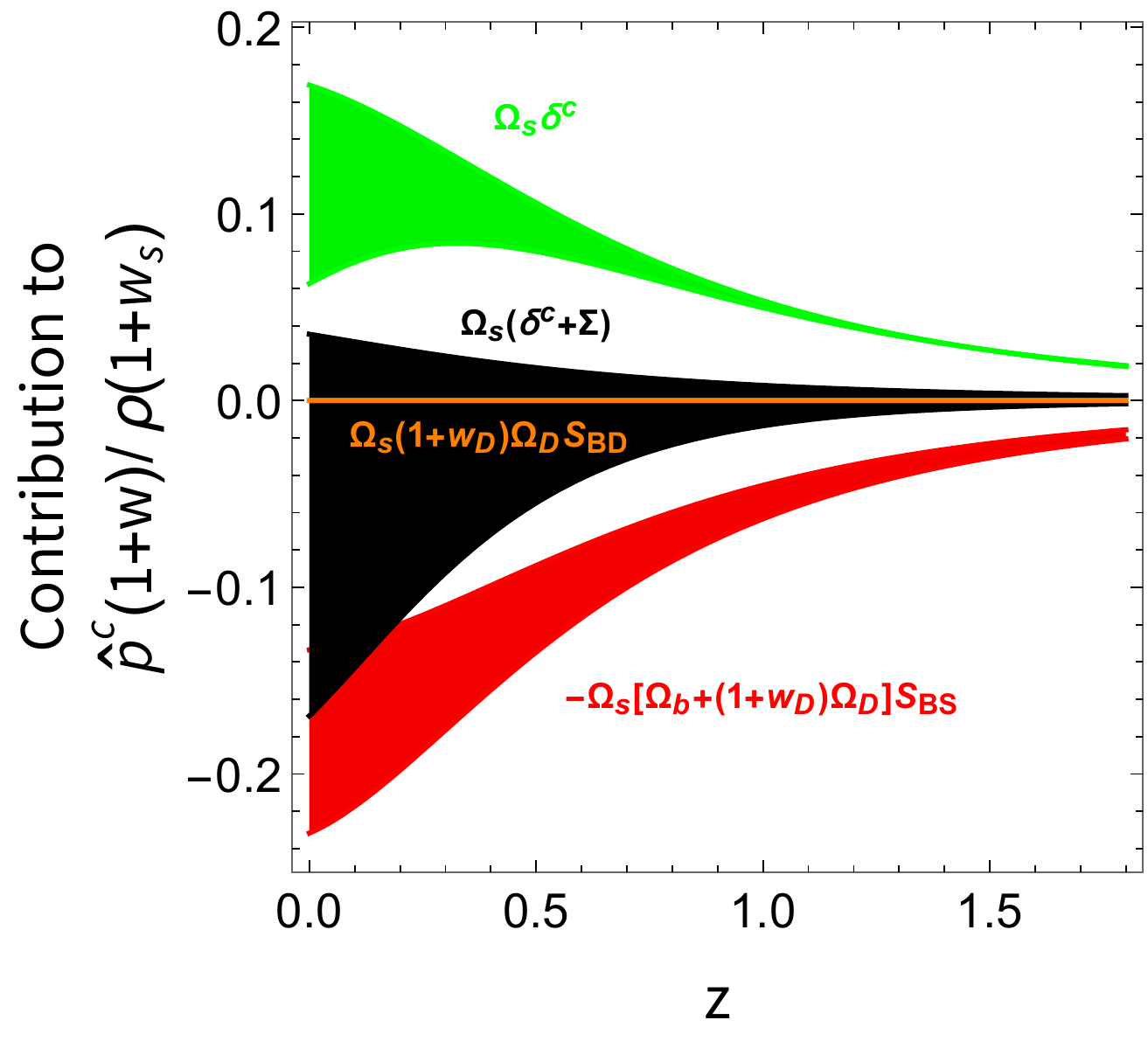}
\caption{Relative contributions of the different terms in Eq. (\ref{hatpfin})  to the total pressure perturbation. Stripes have been obtained with the same parameter value range as in the light blue region of Fig. 1. The total pressure perturbation is given by the black stripe which is the sum of the adiabatic contribution (green) and the non-adiabatic contribution due to $S_{BS}$ (red). The $S_{BD}$ contribution (orange) vanishes. All contributions vanish  asymptotically at high redshifts.}
\label{FigDeltaP}
\end{figure}

It is obvious from the Planck 2018 CPLCDM results (green region in Fig.~1) that upon neglecting non-adiabatic DE perturbations variations in $w_S$ up to the order of $10^{-1}$ seem to be permitted. This is no longer the case, however, if relative entropic DE perturbations are taken into account.  Tiny deviations of the order of $\pm 10^{-3}$ from the cosmological constant behavior $w_{S}=-1$ lead to unacceptably large consequences for the growth rate. This is the main result of the present  work, equivalent to a strong support for the standard $\Lambda$CDM model. 
The necessary accuracy in measuring $w_S$ demands a much larger Figure of Merit in comparison to  what current cosmological surveys can reach. 
The individual contributions of the quantities $\delta^c$, $S_{BD}$ and $S_{BS}$ to the total pressure perturbation are shown in Fig. \ref{FigDeltaP} using the same parameter value range as in the light blue region of Fig. \ref{Fig2}. The black stripe corresponds to the total pressure perturbation i.e., the sum of the adiabatic (green) and the non-adiabatic terms $S_{BD}$ (orange) and $S_{BS}$ (red). Since baryons and DM have been considered pressureless, $S_{BD}$ vanishes.


The obtained limits on $w_S$ are a consequence of properly considering both adiabatic and 
non-adiabatic pressure perturbations. This demonstrates the importance of taking into account 
relative perturbations in a theory of cosmic structure formation, even though DE perturbations 
$\delta_S$ are (almost) negligible.  
Indeed, while $S_{BS}$ turns out to be a growing function of the scale factor for $z\gtrsim 1$, the 
DE perturbations $\delta_S$ remain always small 
since, from (\ref{Relpert2}), $\delta_S \sim (1+w_S) \lesssim 10^{-3}$. 
This is consistent with the well-known result that DE perturbations in dynamical quintessence models remain small for an effective sound speed of the order of 1 \cite{Ratra:1987rm}. 

Via the derivative of the energy-density perturbations in $f\sigma_8(z)$
these pressure perturbations directly influence the matter growth as shown in Fig. 1. 
Our analysis reveals that potential pressure perturbations due to dynamical DE 
leave an imprint on structure formation which seems to be incompatible with current growth-rate data.

{\bf Summary} - 
We established the set of equations (\ref{SBCpr}), (\ref{SBSpr}) and (\ref{dddeltaf}) to study 
the impact of adiabatic and non-adiabatic pressure perturbations on matter clustering. Our non-interacting three-component model assumes an effective DE sound speed equal to one as well as pressure-free baryon and DM components and negligible anisotropic stresses.
We clarified the crucial role of relative perturbations in the cosmic multi-fluid system.  
Quantifying the consequences of potential pressure perturbations for the matter growth rate 
allowed us to strongly constrain the DE-EoS under the given conditions. 
An acceptable matter growth rate is achieved only for $  \vert 1+w_S \vert \lesssim 10^{-4}$. 
In practice this rules out \textbf{a large class of} quintessence type
dynamical DE models with $w_S \neq -1$. {\bf Note added}: While finalizing this work we became aware of the study \cite{Arjona:2020yum} which corroborates our results.

{\bf Acknowledgments}
WZ acknowledges Conselho Nacional de Desenvolvimento Científico e Tecnológico (CNPq) for financial support under project APV 451916/2019-0. JF and HV thank CNPq and Funda\c c\~ao de Amparo \`a Pesquisa e Inova\c c\~ação do Esp\'irito Santo (FAPES) for partial financial support. HV also acknowledges partial support from PROPP/UFOP. RH was supported by Proyecto VRIEA-PUCV 039.449/2020. We acknowledge fruitful discussions with Rodrigo vom Marttens.

\end{document}